\RequirePackage{lineno}
\documentclass[aps,prb,preprint,superscriptaddress]{revtex4}

\usepackage{graphicx}
\usepackage{bm}
\usepackage{amsmath}
\usepackage{amssymb}
\usepackage{hyperref}
\usepackage[version=3]{mhchem}
\usepackage{float}

\begin{document}

\title {Optimizing phonon scattering by nanoprecipitates in lead chalcogenides}

\author {Xiaolong Yang}
\affiliation{Frontier Institute of Science and Technology, and State Key Laboratory for Mechanical Behavior of Materials, Xi'an Jiaotong University, 710054, Xi'an, P. R. China.}

\author {Jes\'us Carrete}
\affiliation{LITEN, CEA-Grenoble, 17 rue des Martyrs, 38054 Grenoble Cedex 9, France}

\author {Zhao Wang}
\email{wzzhao@yahoo.fr}
\affiliation{Frontier Institute of Science and Technology, and State Key Laboratory for Mechanical Behavior of Materials, Xi'an Jiaotong University, 710054, Xi'an, P. R. China.}

\begin{abstract}
We calculate the thermal conductivity of PbTe and PbS with seven different types of nano-precipitates using an ab-initio-based Boltzmann transport approach. We find that precipitates with realistic size distributions can reduce the thermal conductivity well below the predictions of theoretical models assuming a single precipitate size. We explore the question of how to tune this distribution to reduce the thermal conductivity even further. The predicted minimum value is strongly correlated with the phonon spectrum of the host material and with the mass difference between the host and the inclusions.
\end{abstract}


\maketitle

Precise control of heat transport is a fundamental challenge in technological areas such as semiconductor lasers design, phase-change memory development, thermoelectricity or optoelectronics.\cite{Cahill2003,Cahill2014} In particular, thermoelectric devices able to extract electric power from thermal gradients are of widespread interest for applications in energy harvesting and interconnection technologies.\cite{Zebarjadi2012,Vineis2010} The efficiency of such devices hinges on finding materials with low thermal conductivities. Given that nanostructuring enables dramatic reductions of this variable,\cite{Liu2012,Minnich2009} numerous strategies to enhance thermoelectric efficiency by nanostructuring bulk thermoelectrics have been proposed, including substitutional doping, grain boundaries and precipitates.\cite{Snyder2008,Zhao2013,Kanatzidis2010} Among those, nanometer-sized precipitates (or simply nanoprecipitates) in bulk materials are particularly attractive due to their intrinsic nature, which has the potential to enable simpler manufacturing.\cite{Kanatzidis2010,Minnich2009} The effect of nanoprecipitates on thermal conduction has thus become a focus of attention in the search of efficient nano-structured thermoelectric materials.\cite{Biswas2011,He2010,Ohta2012,Ahn2013,Biswas2012}

The physical basis of phonon scattering by nano-particles is well documented in the literature.\cite{Kim2006,Mingo2009,TURK1974,Heremans2005,Ma2008,Pei2011,Katsuyama1998,Zhao2013} When interpreting experimentally measured thermal conductivity data, researchers most often resort to a Matthiessen approach based on the phonon relaxation-time approximation \cite{Mingo2009,MAJUMDAR1993,KLEMENS1955,TURK1974,YING1956,Wang2010,Wang2011} assuming a single average (or effective) nanoparticle size.\cite{Biswas2011,He2010,Ohta2012,Ahn2013,Biswas2012} However, recent measurements have found that in general the size of nanoprecipitates is distributed across a certain length scale ranging from $\sim 0.5$ to $\sim 20\,\mathrm{nm}$.\cite{Biswas2011,He2010,Ohta2012,Ahn2013,Biswas2012,Tan2014} Given the fact that hierarchically architectured nanostructures scatter phonons more efficiently than monodisperse ones,\cite{Biswas2012} the underestimation of the effect of nanoparticles in previous studies is worth exploring. It must be noted that a very recent theoretical work showed that strongly concentrated, bimodal particle size distributions could lower the lattice thermal conductivity of SiGe beyond the single-size limit;\cite{Zhang2015} however, this kind of distribution bears little resemblance to the ones measured from experiments.\cite{Biswas2011,He2010,Ohta2012,Ahn2013,Biswas2012,Tan2014}

With the above motivation, it is crucial to quantitatively understand the effect of experimentally observed precipitate size distributions on the thermal transport properties of thermoelectric composites with a view to enhancing thermoelectric performance. Here we look into this issue by calculating the thermal conductivity of two typical thermoelectric materials -- PbTe and PbS -- with different precipitate types reported in the experimental literature, including \ce{Pb}, \ce{Bi}, \ce{Bi2Te3}, \ce{Sb}, \ce{Na2Te}, \ce{Ag2Te} and \ce{BaTe}, and considering experimentally observable precipitate size distributions.

The lattice thermal conductivity $\kappa$ can be obtained by summing up the contributions from each phonon branch $\alpha$ and integrating over the first Brillouin zone,

\begin{equation}
\label{eq:1}
\kappa=\frac{1}{k_B T^2}\frac{1}{8\pi^3}\sum\limits_{\alpha}\int\limits_{BZ} f_0\left(\omega_{\alpha,q}\right)\left[f_0\left(\omega_{\alpha,q}\right)+1\right]\upsilon_{\alpha,q}^2\hbar^2\omega_{\alpha,q}^2\tau_{\alpha,q} dq,
\end{equation}

\noindent $\upsilon_{\alpha,q}$ the phonon group velocity in the transport direction, $\tau_{\alpha,q}$ the phonon relaxation time and $f_{0}$ denotes the Bose-Einstein distribution. We obtained all those elements for bulk PbTe and PbS from \textit{ab-initio} calculations. We started by relaxing the unit cell parameter of its cubic unit cell using the VASP DFT package,\cite{VASP1,VASP2} with projector-augmented wave datasets,\cite{BLOCHL1994} a $16\times 16\times 16$ Monkhorst-Pack grid in reciprocal space, and under the generalized gradient approximation (GGA).\cite{PBE} We then obtained sets of second- and third-order interatomic force constants from a finite-displacement scheme, employing the open-source software packages Phonopy\cite{Togo2008} and Thirdorder,\cite{Li2014} respectively, and a $4\times 4\times 4$ supercell. Both pieces of software harness the space-group symmetries of PbTe and PbS to greatly reduce the required number of DFT calculations. Those calculations were also performed with VASP using a $4\times 4\times 4$ Monkhorst-Pack grid. Finally, the equilibrium atomic structure and both sets of interatomic force constants were used as inputs to ShengBTE,\cite{Li2014} a solver of the Boltzmann transport equation\cite{Wang2011} for phonons, to obtain all required frequencies, group velocities and relaxation times for the bulk. A comparison between our calculated phonon dispersion and that in the literature, as well as other parameters used in this computation can be found in the Supplemental Material.\cite{a}

When precipitates are added to bulk PbTe and PbS, the total relaxation time $\tau$ can be approximated as a Matthiessen sum of intrinsic (i) and nanoparticle (np) contributions to scattering:

\begin{equation}
\label{eq:5}
\frac{1}{\tau}=\frac{1}{\tau^i}+\frac{1}{\tau^{np}}.
\end{equation}

The intrinsic term, as computed \textit{ab initio} by ShengBTE, in turn includes contributions from all allowed three-phonon processes and from elastic isotopic scattering:

\begin{equation}
  \tau_{\alpha,q}^{i} =
\left[
  \sum\limits^{+}_{\alpha'q'\alpha''q''}\Gamma_{\alpha q\alpha'q'\alpha''q''}^{+}+
  \sum\limits^{-}_{\alpha'q'\alpha''q''}\frac{1}{2}\Gamma_{\alpha q\alpha'q'\alpha''q''}^{-}+
  \sum\limits_{\alpha'q'}\Gamma_{\lambda\lambda^{'}}
\right]^{-1}.
\label{eq:2}
\end{equation}

Here, the $+$ and $-$ superscripts denote sums over allowed absorption and emission processes, respectively, under the constraints of conservation of momentum and energy. Full expressions for all terms are not included here for brevity; readers are referred to Ref.\cite{Li2014} for details.

The precipitate scattering term $\tau^{np}$ is computed by incorporating experimentally measured precipitate size data into the empirical model proposed by Kim and Majumdar,\cite{Kim2006} based on an expression for the scattering cross-section $\sigma$ as an interpolation between a $\propto \omega^4$ Rayleigh-like regime and the frequency-independent geometric limit:\cite{MAJUMDAR1993,YING1956}

\begin{align}
  \sigma_{l}&=\frac{\pi D^2}{9}\left(\frac{\Delta \rho}{\rho}\right)^2\left(\frac{\omega D}{2 \upsilon}\right)^4,\label{eq:7}\\
  \sigma_{s}&=\frac{\pi D^2}{2},\label{eq:6}
\end{align}

\noindent where $l$ and $s$ denote the long and short wavelength limits, respectively, $\Delta \rho$ is the difference between the mass density of the filler and that of the matrix material, $\upsilon$ is the phonon group velocity, and $D$ is diameter of spherical precipitate. Noted that the precipitate scattering could contain contributions from strains and dislocations around the matrix/precipitate interfaces,\cite{He2010,He2010(c)} while the calculations of these contributions could however be hard since they require explicit information from experimental measurements such as the misfit between the sphere and the matrix and the Burgers vector of the dislocation.

In previous models considering only a single effective precipitate size,\cite{Mingo2009,Biswas2011,He2010,Ohta2012,Ahn2013,Biswas2012} the total phonon scattering cross-section $\sigma$ was simply taken as

\begin{equation}
\label{eq:8}
\sigma^{-1}=\sigma_{s}^{-1}+\sigma_{l}^{-1}.
\end{equation}

\noindent We take instead a Gamma distribution\cite{JAMBUNATHAN1954,Kim2006} characterized by a probability density function

\begin{equation}
\label{eq:10a}
f(D) =\frac{D^{A-1}e^{-D/B}}{\Gamma(A)B^{A}},
\end{equation}

\noindent where $A$ is a shape parameter, $B$ is a scale parameter, and $\Gamma$ the usual generalized factorial

\begin{equation}
\label{eq:10b}
\Gamma(A)=\int_0^{\infty}t^{A-1}e^{-t}dt.
\end{equation}

Under this assumption, the contribution of all possible positive precipitate sizes must be taken into account, and Eq.\eqref{eq:8} must be generalized as\cite{Kim2006}

\begin{equation}
\label{eq:9}
\sigma=\int_{0}^{\infty}(\sigma_{s}^{-1}+\sigma_{l}^{-1})^{-1}f(D) dD.
\end{equation}

 Gamma probability function is chosen because 1) It describes positive definite variables; 2) It can describe both exponential and Gaussian distributions as particular cases; 3) It can also describe ``fat tails" (power laws). Knowledge of $\sigma$ is enough to determine the nanoprecipitate contribution to each scattering rate as well as the scattering efficiency $\epsilon$, defined as the ratio between the scattering cross section and the projected surface area of the spherical nanoprecipitate,\cite{Kim2006}

\begin{align}
  \epsilon&=\frac{4\sigma}{\pi D^2},\label{eq:9a}\\
  \frac{1}{\tau^{np}}&=N\upsilon \sigma.\label{eq:10(1)}
\end{align}

The two parameters of $f(D)$ are fitted to experimental data by using the maximum likelihood method. It is important to note that the mean diameter $D_{m}$ of the Gamma distribution and its standard deviation can be written as

\begin{align}
  D_{m} &= AB,\label{eq:10c}\\
  \eta & =\sqrt{A}B,
\end{align}

\noindent and that the precipitate volumetric fraction ($F_{V}$) is related to the precipitate number density $N$ and size distribution as

\begin{equation}
\label{eq:11}
F_{V}=N \int_0^{\infty} \frac{1}{6}\pi D^{3} f(D) dD.
\end{equation}

\begin{figure}[thp]
\centerline{\includegraphics[width=16cm]{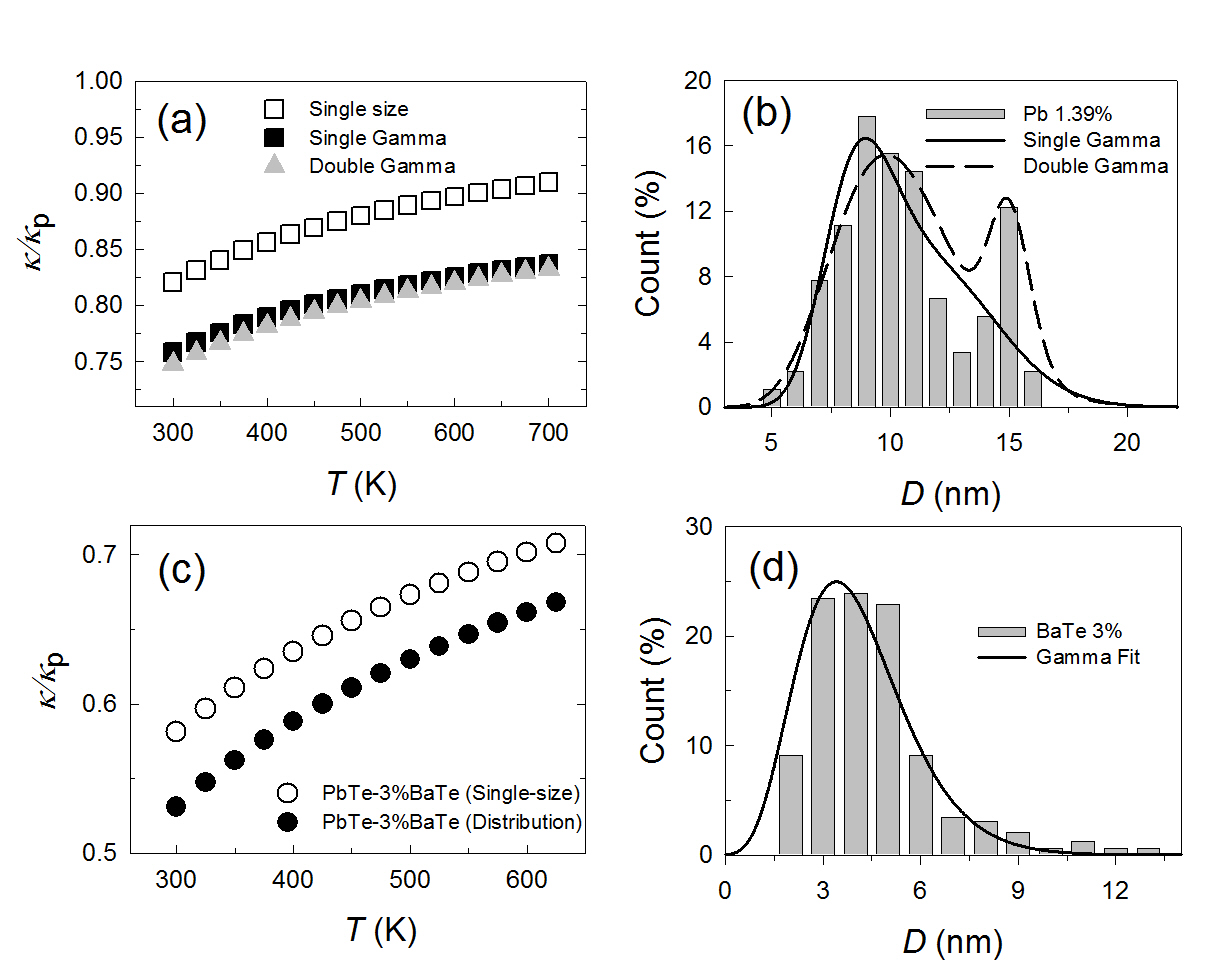}}
\caption{\label{fig1}
(a) and (c) Relative thermal conductivity $\kappa/\kappa_{p}$ ($\kappa_{p}$ is the lattice thermal conductivity of pristine PbTe) \textit{vs.} temperature for PbTe samples with Pb and BaTe nanoprecipitates. Results obtained with  Gamma distribution (filled symbols) are compared to those computed assuming a single average precipitate size (open symbols). (b) and (d) experimentally measured precipitate size distributions of Pb and BaTe nanoprecipitates in PbTe.}
\end{figure}

We start by computing the thermal conductivities of PbTe samples with Pb and BaTe nanoprecipitates, using the experimentally measured precipitate size distributions shown in Fig.\ref{fig1} (b and d).\cite{Lo2012} The results are plotted in Fig.\ref{fig1} (a and c). To take the possibility of multimodal distributions into account, we fit the precipitate size distribution in Fig.\ref{fig1}(b) using a linear combination of two Gamma functions, and compare the results with those given by the single Gamma distribution. Although there is a slight difference in the result, we note that for it to be really significant the importance of the second peak would have to be much larger. It is readily seen that the thermal conductivity predicted by considering a distribution of precipitate sizes is far lower than that given by the single-size approximation. This suggests that phonon scattering by nanoprecipitates in thermoelectric composites can be considerably stronger than previously thought.

\begin{figure}[thp]
\centerline{\includegraphics[width=9cm]{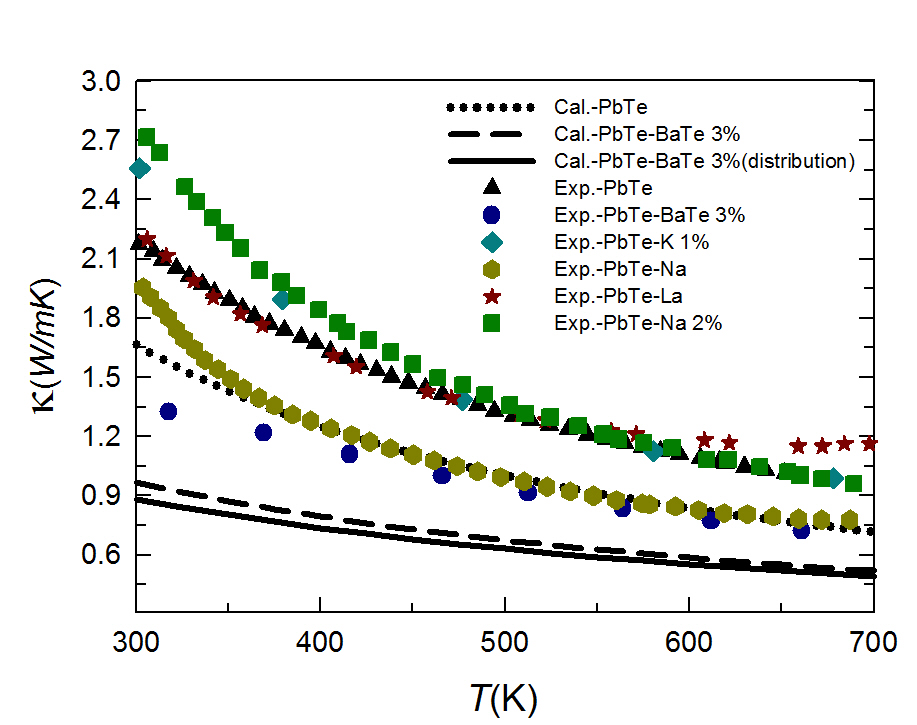}}
\caption{\label{fig2}
Temperature dependence of lattice thermal conductivity from our calculation and experimentally measured results.}
\end{figure}

We compare our calculated thermal conductivity of PbTe with BaTe precipitates with experimental PbTe thermal conductivity data available in the literature (Exp.-PbTe and Exp.-PbTe-BaTe 3\%;\cite{Lo2012} Exp.-PbTe-K 1\%;\cite{Zhang2012} Exp.-PbTe-Na;\cite{Pei2011a} Exp.-PbTe-La;\cite{Pei2014} Exp.-PbTe-Na 2\% \cite{Pei2012})in Fig.\ref{fig2}. It can be seen that our $\kappa$ values are lower than the experimental ones. This may be due to the fact that we use the experimentally measured precipitate distributions in Refs.\cite{Lo2012} as the basic inputs of our calculation, while the experimental statistics may preferably be made in the sample locations with high precipitate density.\cite{Wu2014} It should be noted for this comparison that the experimentally measured thermal conductivities always include multiple contributions from different scattering sources, and it is hard to isolate the influence of precipitate scattering. It is in this context that theoretical approaches to the problem become the most valuable, especially in light of the lack of experimental microscopic information such as grain and precipitate size distributions.

\begin{figure}[thp]
\centerline{\includegraphics[width=8.5cm]{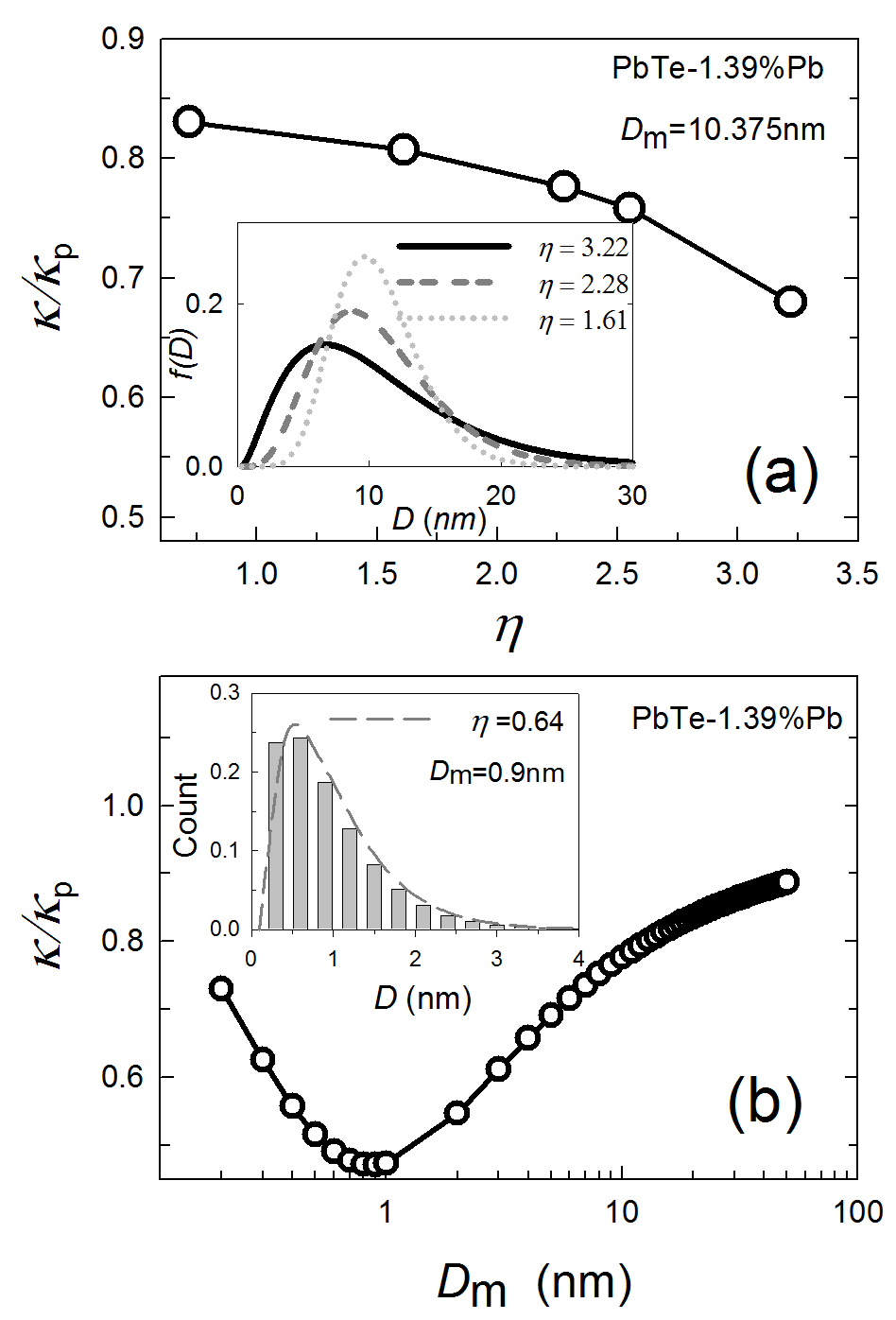}}
\caption{\label{fig3}
(a) Relative thermal conductivity \textit{vs.} standard deviations of the size distribution ($\eta$) at $300\,\mathrm{K}$ for a given average size $D_{m}$. (b) Relative thermal conductivity \textit{vs.} average size $D_{m}$ for a given distribution scale parameter $B=1.0$ at $300\,\mathrm{K}$}
\end{figure}

The results above stem from the fact that precipitates with different sizes enable scattering of a broader part of the phonon spectrum than a single size.\cite{Biswas2012,Zhang2015} This implies that the thermal conductivity must change with the shape of the size distribution. To illustrate this effect, we compute the relative thermal conductivity for distributions with a given average size $D_{m}$ and diverse standard deviations $\eta$ (which describes the breadth of the distribution). Fig.\ref{fig3} (a) shows how the thermal conductivity decreases with increasing $\eta$, as expected. This happens because phonon scattering by nanoprecipitates is gradually enhanced when the distribution is made broader.

Fig.\ref{fig3} (b) shows the thermal conductivity as a function of average size $D_{m}$ for a fixed scale parameter (Eq.\ref{eq:10a}). We observe that the dependence of thermal conductivity on $D_{m}$ is similar to its dependence on particle size in the single-size approximation, as shown in previous work.\cite{Mingo2009} Specifically, an optimal precipitate size can still be found. However, the choice of $D_m$ does not exhaust the degrees of freedom in the problem, and thus there should exist an optimal size distribution that minimizes the thermal conductivity beyond the single-size limit.

\begin{figure}[thp]
\centerline{\includegraphics[width=16cm]{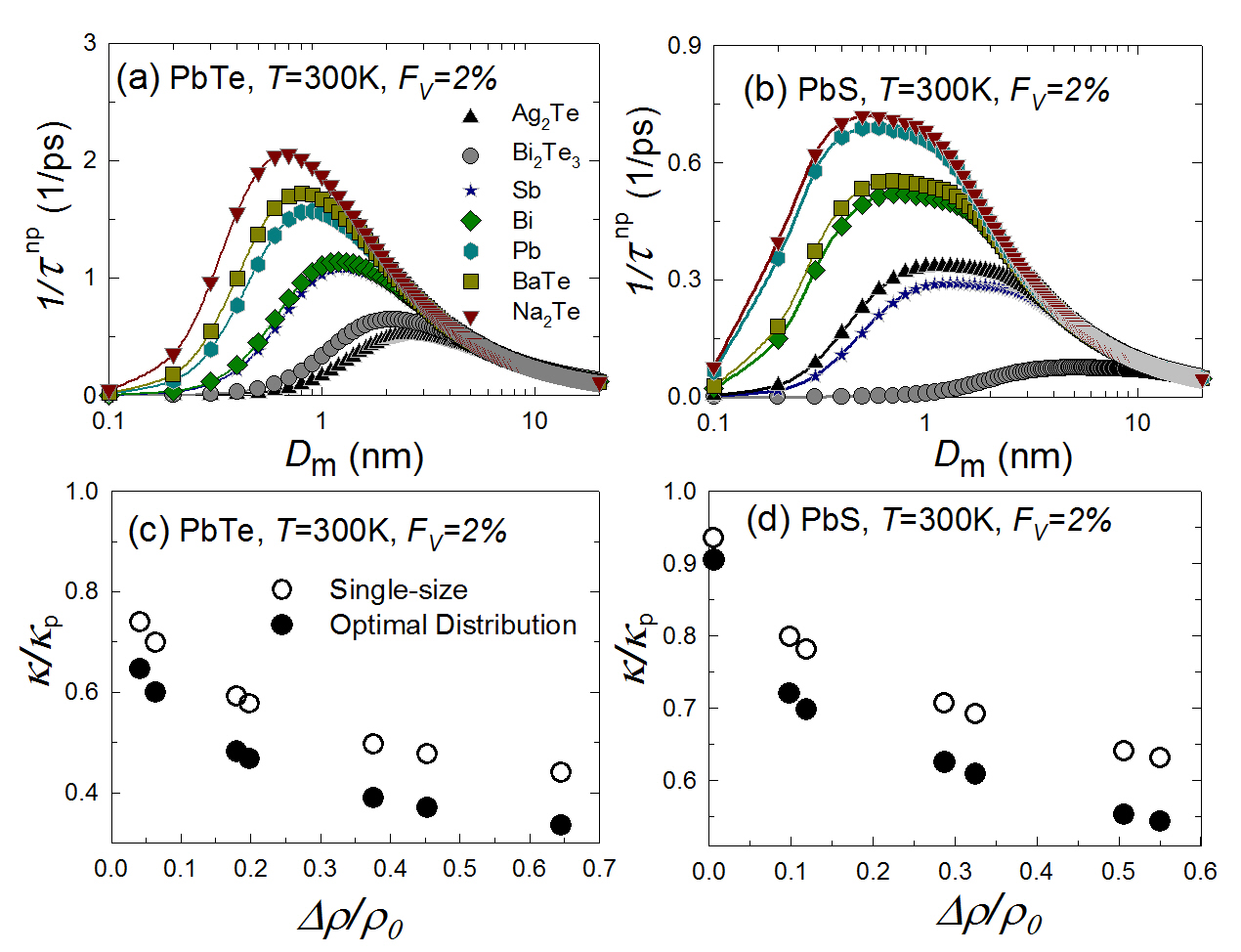}}
\caption{\label{fig4}
(a) and (b) Scattering rate by precipitates $1/\tau^{np}$ as a function of average precipitate size $D_{m}$ for seven different precipitate materials in PbTe and PbS at a given precipitate volume fraction $F_{v}$ at $300\,\mathrm{K}$. (c) and (d) Optimal relative thermal conductivity \textit{vs.} the mass density difference $\Delta D$ between the host and precipitate materials at $300\,\mathrm{K}$. The data represented by the solid symbols correspond to the peaks of curves in (a) and (b).}
\end{figure}

For thermoelectrics, finding that size distribution $f_{opt}(D)$ that reduces the thermal conductivity most efficiently is crucial. We perform a detailed exploration of the $\left(A, B\right)$ space in the search of a maximum precipitate scattering rate $1/\tau^{np}$, a key parameter for estimating the thermal conductivity reduction induced by nano-inclusions [Eq.\eqref{eq:9}] in the present model]. We plot the maximum $1/\tau^{np}$ in Fig.\ref{fig4} (a,b) as a function of average size $D_{m}$ for seven different precipitate materials in PbTe and PbS at a given precipitate volume fraction. The peaks of these curves showing the maxima of $1/\tau^{np}$ corresponding to the lowest thermal conductivity at this precipitate volume fraction. We see the optimal size distribution changes with both the host and the inclusion materials. Based on the dependence of phonon scattering by nanoprecipitates considered here [Eq.\eqref{eq:10(1)}], in Fig.\ref{fig4} (c,d) we plot the optimal thermal conductivity as a function of the mass density difference $\Delta \rho$ between the host and inclusion materials. The optimal values of the relative thermal conductivity are found to be inversely proportional to the mass density difference.

\begin{figure}[thp]
\centerline{\includegraphics[width=16cm]{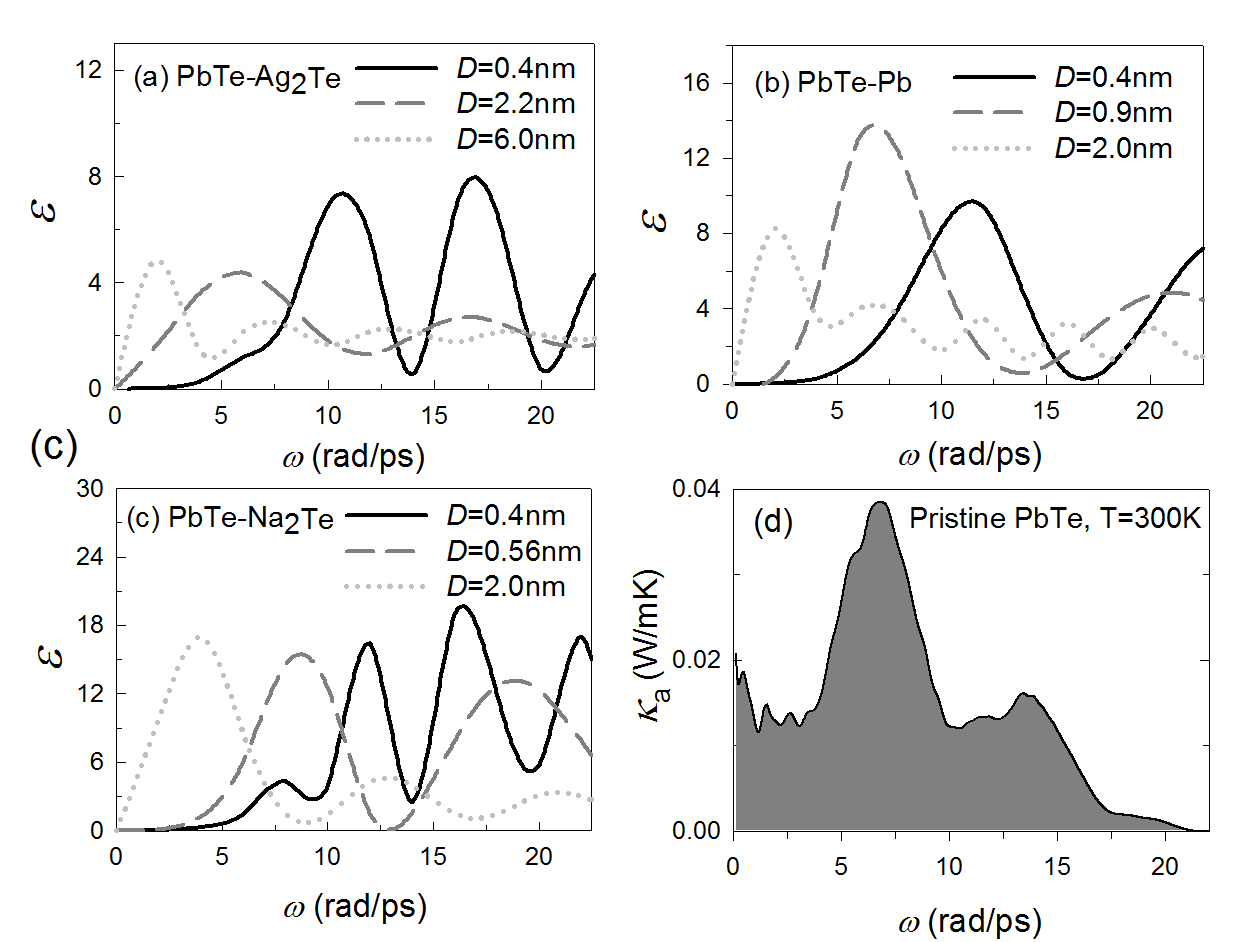}}
\caption{\label{fig5}
(a), (b) and (c) Scattering efficiency of Ag$_{2}$Te, Pb and Na$_{2}$Te nanoprecipitates of different sizes in PbTe as a function of phonon frequency. (d) Contribution to the total thermal conductivity $\kappa_{a}$ from phonons of each frequency for pristine PbTe at $300\,\mathrm{K}$.}
\end{figure}

To further understand the effect of size distribution on thermal conductivity, we plot the scattering efficiency $\epsilon$ [defined in Eq.\eqref{eq:9a}] for several different precipitate diameters as a function of phonon frequency in Fig.\ref{fig5} (a-c), for three different precipitate materials embedded in PbTe. The figure shows that precipitates of different size scatter phonons in different frequency ranges. Taking the contribution to thermal conductivity of each phonon frequency [Fig.\ref{fig5} (d)] as a reference, we see that the combination of several different precipitate sizes covers most of the frequency range where phonons make a major contribution to the thermal conductivity.

Understanding the effect of nanoprecipitates on phonon transport is a crucial step towards the precise control of heat flow in solids. Our results explain the experimental fact that experimentally observed nanoprecipitate distributions lead to decreases in the thermal conductivity beyond the previously-reported single-size limit. We show that the thermal conductivity decreases when increasing the standard deviation of a Gamma distribution of precipitate sizes, and predict the existence of one such distribution that minimizes the thermal conductivity. The parameters of that optimal distribution depend on the mass density difference between the host and the precipitate materials. The characteristic phonon spectrum of the host material is found to be the key for selecting the size range of nanoprecipitates.

\section*{Acknowledgements}
This work is supported by the National Natural Science Foundation of China (Grant No.11204228), the National Basic Research Program of China (2012CB619402 and 2014CB644003), a grant-in-aid of 985 Project from Xi'an Jiaotong University and the Fundamental Research Funds for the Central Universities. We thank Dr. N. Mingo at CEA Grenoble and Prof. J. Li at MIT for helpful discussions.


\begin{thebibliography}{43}
	\expandafter\ifx\csname natexlab\endcsname\relax\def\natexlab#1{#1}\fi
	\expandafter\ifx\csname bibnamefont\endcsname\relax
	\def\bibnamefont#1{#1}\fi
	\expandafter\ifx\csname bibfnamefont\endcsname\relax
	\def\bibfnamefont#1{#1}\fi
	\expandafter\ifx\csname citenamefont\endcsname\relax
	\def\citenamefont#1{#1}\fi
	\expandafter\ifx\csname url\endcsname\relax
	\def\url#1{\texttt{#1}}\fi
	\expandafter\ifx\csname urlprefix\endcsname\relax\def\urlprefix{URL }\fi
	\providecommand{\bibinfo}[2]{#2}
	\providecommand{\eprint}[2][]{\url{#2}}
	
	\bibitem[{\citenamefont{Cahill et~al.}(2003)\citenamefont{Cahill, Ford,
			Goodson, Mahan, Majumdar, Maris, Merlin, and Phillpot}}]{Cahill2003}
	\bibinfo{author}{\bibfnamefont{D.~G.} \bibnamefont{Cahill}},
	\bibinfo{author}{\bibfnamefont{W.~K.} \bibnamefont{Ford}},
	\bibinfo{author}{\bibfnamefont{K.~E.} \bibnamefont{Goodson}},
	\bibinfo{author}{\bibfnamefont{G.~D.} \bibnamefont{Mahan}},
	\bibinfo{author}{\bibfnamefont{A.}~\bibnamefont{Majumdar}},
	\bibinfo{author}{\bibfnamefont{H.~J.} \bibnamefont{Maris}},
	\bibinfo{author}{\bibfnamefont{R.}~\bibnamefont{Merlin}}, \bibnamefont{and}
	\bibinfo{author}{\bibfnamefont{S.~R.} \bibnamefont{Phillpot}},
	\bibinfo{journal}{J. Appl. Phys.} \textbf{\bibinfo{volume}{93}},
	\bibinfo{pages}{793} (\bibinfo{year}{2003}).
	
	\bibitem[{\citenamefont{Cahill et~al.}(2014)\citenamefont{Cahill, Braun, Chen,
			Clarke, Fan, Goodson, Keblinski, King, Mahan, Majumdar et~al.}}]{Cahill2014}
	\bibinfo{author}{\bibfnamefont{D.~G.} \bibnamefont{Cahill}},
	\bibinfo{author}{\bibfnamefont{P.~V.} \bibnamefont{Braun}},
	\bibinfo{author}{\bibfnamefont{G.}~\bibnamefont{Chen}},
	\bibinfo{author}{\bibfnamefont{D.~R.} \bibnamefont{Clarke}},
	\bibinfo{author}{\bibfnamefont{S.}~\bibnamefont{Fan}},
	\bibinfo{author}{\bibfnamefont{K.~E.} \bibnamefont{Goodson}},
	\bibinfo{author}{\bibfnamefont{P.}~\bibnamefont{Keblinski}},
	\bibinfo{author}{\bibfnamefont{W.~P.} \bibnamefont{King}},
	\bibinfo{author}{\bibfnamefont{G.~D.} \bibnamefont{Mahan}},
	\bibinfo{author}{\bibfnamefont{A.}~\bibnamefont{Majumdar}},
	\bibnamefont{et~al.}, \bibinfo{journal}{Appl. Phys. Rev.}
	\textbf{\bibinfo{volume}{1}}, \bibinfo{pages}{011305} (\bibinfo{year}{2014}).
	
	\bibitem[{\citenamefont{Zebarjadi et~al.}(2012)\citenamefont{Zebarjadi,
			Esfarjani, Dresselhaus, Ren, and Chen}}]{Zebarjadi2012}
	\bibinfo{author}{\bibfnamefont{M.}~\bibnamefont{Zebarjadi}},
	\bibinfo{author}{\bibfnamefont{K.}~\bibnamefont{Esfarjani}},
	\bibinfo{author}{\bibfnamefont{M.~S.} \bibnamefont{Dresselhaus}},
	\bibinfo{author}{\bibfnamefont{Z.~F.} \bibnamefont{Ren}}, \bibnamefont{and}
	\bibinfo{author}{\bibfnamefont{G.}~\bibnamefont{Chen}},
	\bibinfo{journal}{Energ. Environ. Sci.} \textbf{\bibinfo{volume}{5}},
	\bibinfo{pages}{5147} (\bibinfo{year}{2012}).
	
	\bibitem[{\citenamefont{Vineis et~al.}(2010)\citenamefont{Vineis, Shakouri,
			Majumdar, and Kanatzidis}}]{Vineis2010}
	\bibinfo{author}{\bibfnamefont{C.~J.} \bibnamefont{Vineis}},
	\bibinfo{author}{\bibfnamefont{A.}~\bibnamefont{Shakouri}},
	\bibinfo{author}{\bibfnamefont{A.}~\bibnamefont{Majumdar}}, \bibnamefont{and}
	\bibinfo{author}{\bibfnamefont{M.~G.} \bibnamefont{Kanatzidis}},
	\bibinfo{journal}{Adv. Mater.} \textbf{\bibinfo{volume}{22}},
	\bibinfo{pages}{3970} (\bibinfo{year}{2010}).
	
	\bibitem[{\citenamefont{Liu et~al.}(2012)\citenamefont{Liu, Yan, Chen, and
			Ren}}]{Liu2012}
	\bibinfo{author}{\bibfnamefont{W.}~\bibnamefont{Liu}},
	\bibinfo{author}{\bibfnamefont{X.}~\bibnamefont{Yan}},
	\bibinfo{author}{\bibfnamefont{G.}~\bibnamefont{Chen}}, \bibnamefont{and}
	\bibinfo{author}{\bibfnamefont{Z.}~\bibnamefont{Ren}}, \bibinfo{journal}{Nano
		Energy} \textbf{\bibinfo{volume}{1}}, \bibinfo{pages}{42}
	(\bibinfo{year}{2012}).
	
	\bibitem[{\citenamefont{Minnich et~al.}(2009)\citenamefont{Minnich,
			Dresselhaus, Ren, and Chen}}]{Minnich2009}
	\bibinfo{author}{\bibfnamefont{A.~J.} \bibnamefont{Minnich}},
	\bibinfo{author}{\bibfnamefont{M.~S.} \bibnamefont{Dresselhaus}},
	\bibinfo{author}{\bibfnamefont{Z.~F.} \bibnamefont{Ren}}, \bibnamefont{and}
	\bibinfo{author}{\bibfnamefont{G.}~\bibnamefont{Chen}},
	\bibinfo{journal}{Energ. Environ. Sci.} \textbf{\bibinfo{volume}{2}},
	\bibinfo{pages}{466} (\bibinfo{year}{2009}).
	
	\bibitem[{\citenamefont{Snyder and Toberer}(2008)}]{Snyder2008}
	\bibinfo{author}{\bibfnamefont{G.~J.} \bibnamefont{Snyder}} \bibnamefont{and}
	\bibinfo{author}{\bibfnamefont{E.~S.} \bibnamefont{Toberer}},
	\bibinfo{journal}{Nat. Mater.} \textbf{\bibinfo{volume}{7}},
	\bibinfo{pages}{105} (\bibinfo{year}{2008}).
	
	\bibitem[{\citenamefont{Zhao et~al.}(2013)\citenamefont{Zhao, Wu, Hao, Wu,
			Zhou, Biswas, He, Hogan, Uher, Wolverton et~al.}}]{Zhao2013}
	\bibinfo{author}{\bibfnamefont{L.~D.} \bibnamefont{Zhao}},
	\bibinfo{author}{\bibfnamefont{H.~J.} \bibnamefont{Wu}},
	\bibinfo{author}{\bibfnamefont{S.~Q.} \bibnamefont{Hao}},
	\bibinfo{author}{\bibfnamefont{C.~I.} \bibnamefont{Wu}},
	\bibinfo{author}{\bibfnamefont{X.~Y.} \bibnamefont{Zhou}},
	\bibinfo{author}{\bibfnamefont{K.}~\bibnamefont{Biswas}},
	\bibinfo{author}{\bibfnamefont{J.~Q.} \bibnamefont{He}},
	\bibinfo{author}{\bibfnamefont{T.~P.} \bibnamefont{Hogan}},
	\bibinfo{author}{\bibfnamefont{C.}~\bibnamefont{Uher}},
	\bibinfo{author}{\bibfnamefont{C.}~\bibnamefont{Wolverton}},
	\bibnamefont{et~al.}, \bibinfo{journal}{Energ. Environ. Sci.}
	\textbf{\bibinfo{volume}{6}}, \bibinfo{pages}{3346} (\bibinfo{year}{2013}).
	
	\bibitem[{\citenamefont{Kanatzidis}(2010)}]{Kanatzidis2010}
	\bibinfo{author}{\bibfnamefont{M.~G.} \bibnamefont{Kanatzidis}},
	\bibinfo{journal}{Chem. Mater.} \textbf{\bibinfo{volume}{22}},
	\bibinfo{pages}{648} (\bibinfo{year}{2010}).
	
	\bibitem[{\citenamefont{Biswas et~al.}(2011)\citenamefont{Biswas, He, Zhang,
			Wang, Uher, Dravid, and Kanatzidis}}]{Biswas2011}
	\bibinfo{author}{\bibfnamefont{K.}~\bibnamefont{Biswas}},
	\bibinfo{author}{\bibfnamefont{J.}~\bibnamefont{He}},
	\bibinfo{author}{\bibfnamefont{Q.}~\bibnamefont{Zhang}},
	\bibinfo{author}{\bibfnamefont{G.}~\bibnamefont{Wang}},
	\bibinfo{author}{\bibfnamefont{C.}~\bibnamefont{Uher}},
	\bibinfo{author}{\bibfnamefont{V.~P.} \bibnamefont{Dravid}},
	\bibnamefont{and} \bibinfo{author}{\bibfnamefont{M.~G.}
		\bibnamefont{Kanatzidis}}, \bibinfo{journal}{Nat. Chem.}
	\textbf{\bibinfo{volume}{3}}, \bibinfo{pages}{160} (\bibinfo{year}{2011}).
	
	\bibitem[{\citenamefont{He et~al.}(2010{\natexlab{a}})\citenamefont{He, Girard,
			Kanatzidis, and Dravid}}]{He2010}
	\bibinfo{author}{\bibfnamefont{J.}~\bibnamefont{He}},
	\bibinfo{author}{\bibfnamefont{S.~N.} \bibnamefont{Girard}},
	\bibinfo{author}{\bibfnamefont{M.~G.} \bibnamefont{Kanatzidis}},
	\bibnamefont{and} \bibinfo{author}{\bibfnamefont{V.~P.}
		\bibnamefont{Dravid}}, \bibinfo{journal}{Adv. Funct. Mater.}
	\textbf{\bibinfo{volume}{20}}, \bibinfo{pages}{764}
	(\bibinfo{year}{2010}{\natexlab{a}}).
	
	\bibitem[{\citenamefont{Ohta et~al.}(2012)\citenamefont{Ohta, Biswas, Lo, He,
			Chung, Dravid, and Kanatzidis}}]{Ohta2012}
	\bibinfo{author}{\bibfnamefont{M.}~\bibnamefont{Ohta}},
	\bibinfo{author}{\bibfnamefont{K.}~\bibnamefont{Biswas}},
	\bibinfo{author}{\bibfnamefont{S.-H.} \bibnamefont{Lo}},
	\bibinfo{author}{\bibfnamefont{J.}~\bibnamefont{He}},
	\bibinfo{author}{\bibfnamefont{D.~Y.} \bibnamefont{Chung}},
	\bibinfo{author}{\bibfnamefont{V.~P.} \bibnamefont{Dravid}},
	\bibnamefont{and} \bibinfo{author}{\bibfnamefont{M.~G.}
		\bibnamefont{Kanatzidis}}, \bibinfo{journal}{Adv. Energy Mater.}
	\textbf{\bibinfo{volume}{2}}, \bibinfo{pages}{1117} (\bibinfo{year}{2012}).
	
	\bibitem[{\citenamefont{Ahn et~al.}(2013)\citenamefont{Ahn, Biswas, He, Chung,
			Dravid, and Kanatzidis}}]{Ahn2013}
	\bibinfo{author}{\bibfnamefont{K.}~\bibnamefont{Ahn}},
	\bibinfo{author}{\bibfnamefont{K.}~\bibnamefont{Biswas}},
	\bibinfo{author}{\bibfnamefont{J.}~\bibnamefont{He}},
	\bibinfo{author}{\bibfnamefont{I.}~\bibnamefont{Chung}},
	\bibinfo{author}{\bibfnamefont{V.}~\bibnamefont{Dravid}}, \bibnamefont{and}
	\bibinfo{author}{\bibfnamefont{M.~G.} \bibnamefont{Kanatzidis}},
	\bibinfo{journal}{Energ. Environ. Sci.} \textbf{\bibinfo{volume}{6}},
	\bibinfo{pages}{1529} (\bibinfo{year}{2013}).
	
	\bibitem[{\citenamefont{Biswas et~al.}(2012)\citenamefont{Biswas, He, Blum, Wu,
			Hogan, Seidman, Dravid, and Kanatzidis}}]{Biswas2012}
	\bibinfo{author}{\bibfnamefont{K.}~\bibnamefont{Biswas}},
	\bibinfo{author}{\bibfnamefont{J.}~\bibnamefont{He}},
	\bibinfo{author}{\bibfnamefont{I.~D.} \bibnamefont{Blum}},
	\bibinfo{author}{\bibfnamefont{C.-I.} \bibnamefont{Wu}},
	\bibinfo{author}{\bibfnamefont{T.~P.} \bibnamefont{Hogan}},
	\bibinfo{author}{\bibfnamefont{D.~N.} \bibnamefont{Seidman}},
	\bibinfo{author}{\bibfnamefont{V.~P.} \bibnamefont{Dravid}},
	\bibnamefont{and} \bibinfo{author}{\bibfnamefont{M.~G.}
		\bibnamefont{Kanatzidis}}, \bibinfo{journal}{Nature}
	\textbf{\bibinfo{volume}{489}}, \bibinfo{pages}{414} (\bibinfo{year}{2012}).
	
	\bibitem[{\citenamefont{Kim and Majumdar}(2006)}]{Kim2006}
	\bibinfo{author}{\bibfnamefont{W.}~\bibnamefont{Kim}} \bibnamefont{and}
	\bibinfo{author}{\bibfnamefont{A.}~\bibnamefont{Majumdar}},
	\bibinfo{journal}{J. Appl. Phys.} \textbf{\bibinfo{volume}{99}},
	\bibinfo{pages}{084306} (\bibinfo{year}{2006}).
	
	\bibitem[{\citenamefont{Mingo et~al.}(2009)\citenamefont{Mingo, Hauser,
			Kobayashi, Plissonnier, and Shakouri}}]{Mingo2009}
	\bibinfo{author}{\bibfnamefont{N.}~\bibnamefont{Mingo}},
	\bibinfo{author}{\bibfnamefont{D.}~\bibnamefont{Hauser}},
	\bibinfo{author}{\bibfnamefont{N.~P.} \bibnamefont{Kobayashi}},
	\bibinfo{author}{\bibfnamefont{M.}~\bibnamefont{Plissonnier}},
	\bibnamefont{and} \bibinfo{author}{\bibfnamefont{A.}~\bibnamefont{Shakouri}},
	\bibinfo{journal}{Nano Lett.} \textbf{\bibinfo{volume}{9}},
	\bibinfo{pages}{711} (\bibinfo{year}{2009}).
	
	\bibitem[{\citenamefont{Turk and Klemens}(1974)}]{TURK1974}
	\bibinfo{author}{\bibfnamefont{L.~A.} \bibnamefont{Turk}} \bibnamefont{and}
	\bibinfo{author}{\bibfnamefont{P.~G.} \bibnamefont{Klemens}},
	\bibinfo{journal}{Phys. Rev. B} \textbf{\bibinfo{volume}{9}},
	\bibinfo{pages}{4422} (\bibinfo{year}{1974}).
	
	\bibitem[{\citenamefont{Heremans et~al.}(2005)\citenamefont{Heremans, Thrush,
			and Morelli}}]{Heremans2005}
	\bibinfo{author}{\bibfnamefont{J.~P.} \bibnamefont{Heremans}},
	\bibinfo{author}{\bibfnamefont{C.~M.} \bibnamefont{Thrush}},
	\bibnamefont{and} \bibinfo{author}{\bibfnamefont{D.~T.}
		\bibnamefont{Morelli}}, \bibinfo{journal}{J. Appl. Phys.}
	\textbf{\bibinfo{volume}{98}}, \bibinfo{pages}{063703}
	(\bibinfo{year}{2005}).
	
	\bibitem[{\citenamefont{Ma et~al.}(2008)\citenamefont{Ma, Hao, Poudel, Lan, Yu,
			Wang, Chen, and Ren}}]{Ma2008}
	\bibinfo{author}{\bibfnamefont{Y.}~\bibnamefont{Ma}},
	\bibinfo{author}{\bibfnamefont{Q.}~\bibnamefont{Hao}},
	\bibinfo{author}{\bibfnamefont{B.}~\bibnamefont{Poudel}},
	\bibinfo{author}{\bibfnamefont{Y.}~\bibnamefont{Lan}},
	\bibinfo{author}{\bibfnamefont{B.}~\bibnamefont{Yu}},
	\bibinfo{author}{\bibfnamefont{D.}~\bibnamefont{Wang}},
	\bibinfo{author}{\bibfnamefont{G.}~\bibnamefont{Chen}}, \bibnamefont{and}
	\bibinfo{author}{\bibfnamefont{Z.}~\bibnamefont{Ren}}, \bibinfo{journal}{Nano
		Lett.} \textbf{\bibinfo{volume}{8}}, \bibinfo{pages}{2580}
	(\bibinfo{year}{2008}).
	
	\bibitem[{\citenamefont{Pei et~al.}(2011{\natexlab{a}})\citenamefont{Pei,
			Lensch-Falk, Toberer, Medlin, and Snyder}}]{Pei2011}
	\bibinfo{author}{\bibfnamefont{Y.}~\bibnamefont{Pei}},
	\bibinfo{author}{\bibfnamefont{J.}~\bibnamefont{Lensch-Falk}},
	\bibinfo{author}{\bibfnamefont{E.~S.} \bibnamefont{Toberer}},
	\bibinfo{author}{\bibfnamefont{D.~L.} \bibnamefont{Medlin}},
	\bibnamefont{and} \bibinfo{author}{\bibfnamefont{G.~J.}
		\bibnamefont{Snyder}}, \bibinfo{journal}{Adv. Funct. Mater.}
	\textbf{\bibinfo{volume}{21}}, \bibinfo{pages}{241}
	(\bibinfo{year}{2011}{\natexlab{a}}).
	
	\bibitem[{\citenamefont{Katsuyama et~al.}(1998)\citenamefont{Katsuyama,
			Shichijo, Ito, Majima, and Nagai}}]{Katsuyama1998}
	\bibinfo{author}{\bibfnamefont{S.}~\bibnamefont{Katsuyama}},
	\bibinfo{author}{\bibfnamefont{Y.}~\bibnamefont{Shichijo}},
	\bibinfo{author}{\bibfnamefont{M.}~\bibnamefont{Ito}},
	\bibinfo{author}{\bibfnamefont{K.}~\bibnamefont{Majima}}, \bibnamefont{and}
	\bibinfo{author}{\bibfnamefont{H.}~\bibnamefont{Nagai}}, \bibinfo{journal}{J.
		Appl. Phys.} \textbf{\bibinfo{volume}{84}}, \bibinfo{pages}{6708}
	(\bibinfo{year}{1998}).
	
	\bibitem[{\citenamefont{Majumdar}(1993)}]{MAJUMDAR1993}
	\bibinfo{author}{\bibfnamefont{A.}~\bibnamefont{Majumdar}},
	\bibinfo{journal}{J. Heat Transfer-trans Asme}
	\textbf{\bibinfo{volume}{115}}, \bibinfo{pages}{7} (\bibinfo{year}{1993}).
	
	\bibitem[{\citenamefont{Klemens}(1955)}]{KLEMENS1955}
	\bibinfo{author}{\bibfnamefont{P.~G.} \bibnamefont{Klemens}},
	\bibinfo{journal}{Proc. Phys. Soc. London Sec A}
	\textbf{\bibinfo{volume}{68}}, \bibinfo{pages}{1113} (\bibinfo{year}{1955}).
	
	\bibitem[{\citenamefont{Ying and Truell}(1956)}]{YING1956}
	\bibinfo{author}{\bibfnamefont{C.~F.} \bibnamefont{Ying}} \bibnamefont{and}
	\bibinfo{author}{\bibfnamefont{R.}~\bibnamefont{Truell}},
	\bibinfo{journal}{Appl. Phys. Lett.} \textbf{\bibinfo{volume}{27}},
	\bibinfo{pages}{1086} (\bibinfo{year}{1956}).
	
	\bibitem[{\citenamefont{Wang and Mingo}(2010)}]{Wang2010}
	\bibinfo{author}{\bibfnamefont{Z.}~\bibnamefont{Wang}} \bibnamefont{and}
	\bibinfo{author}{\bibfnamefont{N.}~\bibnamefont{Mingo}},
	\bibinfo{journal}{Appl. Phys. Lett.} \textbf{\bibinfo{volume}{97}},
	\bibinfo{pages}{101903} (\bibinfo{year}{2010}).
	
	\bibitem[{\citenamefont{Wang and Mingo}(2011)}]{Wang2011}
	\bibinfo{author}{\bibfnamefont{Z.}~\bibnamefont{Wang}} \bibnamefont{and}
	\bibinfo{author}{\bibfnamefont{N.}~\bibnamefont{Mingo}},
	\bibinfo{journal}{Appl. Phys. Lett.} \textbf{\bibinfo{volume}{99}},
	\bibinfo{pages}{101903} (\bibinfo{year}{2011}).
	
	\bibitem[{\citenamefont{Tan et~al.}(2014)\citenamefont{Tan, Zhao, Li, Wu, Wei,
			Xing, and Kanatzidis}}]{Tan2014}
	\bibinfo{author}{\bibfnamefont{Q.}~\bibnamefont{Tan}},
	\bibinfo{author}{\bibfnamefont{L.-D.} \bibnamefont{Zhao}},
	\bibinfo{author}{\bibfnamefont{J.-F.} \bibnamefont{Li}},
	\bibinfo{author}{\bibfnamefont{C.-F.} \bibnamefont{Wu}},
	\bibinfo{author}{\bibfnamefont{T.-R.} \bibnamefont{Wei}},
	\bibinfo{author}{\bibfnamefont{Z.-B.} \bibnamefont{Xing}}, \bibnamefont{and}
	\bibinfo{author}{\bibfnamefont{M.~G.} \bibnamefont{Kanatzidis}},
	\bibinfo{journal}{J. Mater. Chem. A} \textbf{\bibinfo{volume}{2}},
	\bibinfo{pages}{17302} (\bibinfo{year}{2014}).
	
	\bibitem[{\citenamefont{Zhang and Minnich}(2015)}]{Zhang2015}
	\bibinfo{author}{\bibfnamefont{H.}~\bibnamefont{Zhang}} \bibnamefont{and}
	\bibinfo{author}{\bibfnamefont{A.~J.} \bibnamefont{Minnich}},
	\bibinfo{journal}{Sci. R.} \textbf{\bibinfo{volume}{5}},
	\bibinfo{pages}{8995} (\bibinfo{year}{2015}).
	
	\bibitem[{\citenamefont{Kresse and Hafner}(1993)}]{VASP1}
	\bibinfo{author}{\bibfnamefont{G.}~\bibnamefont{Kresse}} \bibnamefont{and}
	\bibinfo{author}{\bibfnamefont{J.}~\bibnamefont{Hafner}},
	\bibinfo{journal}{Phys. Rev. B} \textbf{\bibinfo{volume}{47}},
	\bibinfo{pages}{558} (\bibinfo{year}{1993}).
	
	\bibitem[{\citenamefont{Kresse and Furthm\"uller}(1996)}]{VASP2}
	\bibinfo{author}{\bibfnamefont{G.}~\bibnamefont{Kresse}} \bibnamefont{and}
	\bibinfo{author}{\bibfnamefont{J.}~\bibnamefont{Furthm\"uller}},
	\bibinfo{journal}{Comp. Mat. Sci.} \textbf{\bibinfo{volume}{6}},
	\bibinfo{pages}{15} (\bibinfo{year}{1996}).
	
	\bibitem[{\citenamefont{Blochl}(1994)}]{BLOCHL1994}
	\bibinfo{author}{\bibfnamefont{P.~E.} \bibnamefont{Blochl}},
	\bibinfo{journal}{Phys. Rev.B} \textbf{\bibinfo{volume}{50}},
	\bibinfo{pages}{17953} (\bibinfo{year}{1994}).
	
	\bibitem[{\citenamefont{Perdew et~al.}(1996)\citenamefont{Perdew, Burke, and
			Ernzerhof}}]{PBE}
	\bibinfo{author}{\bibfnamefont{J.}~\bibnamefont{Perdew}},
	\bibinfo{author}{\bibfnamefont{K.}~\bibnamefont{Burke}}, \bibnamefont{and}
	\bibinfo{author}{\bibfnamefont{M.}~\bibnamefont{Ernzerhof}},
	\bibinfo{journal}{Phys. Rev. Lett.} \textbf{\bibinfo{volume}{77}},
	\bibinfo{pages}{3865} (\bibinfo{year}{1996}).
	
	\bibitem[{\citenamefont{Togo et~al.}(2008)\citenamefont{Togo, Oba, and
			Tanaka}}]{Togo2008}
	\bibinfo{author}{\bibfnamefont{A.}~\bibnamefont{Togo}},
	\bibinfo{author}{\bibfnamefont{F.}~\bibnamefont{Oba}}, \bibnamefont{and}
	\bibinfo{author}{\bibfnamefont{I.}~\bibnamefont{Tanaka}},
	\bibinfo{journal}{Phys. Rev. B} \textbf{\bibinfo{volume}{78}},
	\bibinfo{pages}{134106} (\bibinfo{year}{2008}).
	
	\bibitem[{\citenamefont{Li et~al.}(2014)\citenamefont{Li, Carrete, Katcho, and
			Mingo}}]{Li2014}
	\bibinfo{author}{\bibfnamefont{W.}~\bibnamefont{Li}},
	\bibinfo{author}{\bibfnamefont{J.}~\bibnamefont{Carrete}},
	\bibinfo{author}{\bibfnamefont{N.~A.} \bibnamefont{Katcho}},
	\bibnamefont{and} \bibinfo{author}{\bibfnamefont{N.}~\bibnamefont{Mingo}},
	\bibinfo{journal}{Comput. Phys. Commun.} \textbf{\bibinfo{volume}{185}},
	\bibinfo{pages}{1747} (\bibinfo{year}{2014}).
	
	\bibitem[{a()}]{a}
	See supplemental material at \url{ [URL]} for computational details of band dispersion.
	
	\bibitem[{\citenamefont{He et~al.}(2010{\natexlab{b}})\citenamefont{He,
			Sootsman, Girard, Zheng, Wen, Zhu, Kanatzidis, and Dravid}}]{He2010(c)}
	\bibinfo{author}{\bibfnamefont{J.}~\bibnamefont{He}},
	\bibinfo{author}{\bibfnamefont{J.~R.} \bibnamefont{Sootsman}},
	\bibinfo{author}{\bibfnamefont{S.~N.} \bibnamefont{Girard}},
	\bibinfo{author}{\bibfnamefont{J.-C.} \bibnamefont{Zheng}},
	\bibinfo{author}{\bibfnamefont{J.}~\bibnamefont{Wen}},
	\bibinfo{author}{\bibfnamefont{Y.}~\bibnamefont{Zhu}},
	\bibinfo{author}{\bibfnamefont{M.~G.} \bibnamefont{Kanatzidis}},
	\bibnamefont{and} \bibinfo{author}{\bibfnamefont{V.~P.}
		\bibnamefont{Dravid}}, \bibinfo{journal}{J. Am. Chem. Soc.}
	\textbf{\bibinfo{volume}{132}}, \bibinfo{pages}{8669}
	(\bibinfo{year}{2010}{\natexlab{b}}).
	
	\bibitem[{\citenamefont{Jambunathan}(1954)}]{JAMBUNATHAN1954}
	\bibinfo{author}{\bibfnamefont{M.~V.} \bibnamefont{Jambunathan}},
	\bibinfo{journal}{Ann. Math. Stat.} \textbf{\bibinfo{volume}{25}},
	\bibinfo{pages}{401} (\bibinfo{year}{1954}).
	
	\bibitem[{\citenamefont{Lo et~al.}(2012)\citenamefont{Lo, He, Biswas,
			Kanatzidis, and Dravid}}]{Lo2012}
	\bibinfo{author}{\bibfnamefont{S.-H.} \bibnamefont{Lo}},
	\bibinfo{author}{\bibfnamefont{J.}~\bibnamefont{He}},
	\bibinfo{author}{\bibfnamefont{K.}~\bibnamefont{Biswas}},
	\bibinfo{author}{\bibfnamefont{M.~G.} \bibnamefont{Kanatzidis}},
	\bibnamefont{and} \bibinfo{author}{\bibfnamefont{V.~P.}
		\bibnamefont{Dravid}}, \bibinfo{journal}{Adv. Funct. Mater.}
	\textbf{\bibinfo{volume}{22}}, \bibinfo{pages}{5175} (\bibinfo{year}{2012}).
	
	\bibitem[{\citenamefont{Zhang et~al.}(2012)\citenamefont{Zhang, Cao, Liu,
			Lukas, Yu, Chen, Opeil, Broido, Chen, and Ren}}]{Zhang2012}
	\bibinfo{author}{\bibfnamefont{Q.}~\bibnamefont{Zhang}},
	\bibinfo{author}{\bibfnamefont{F.}~\bibnamefont{Cao}},
	\bibinfo{author}{\bibfnamefont{W.}~\bibnamefont{Liu}},
	\bibinfo{author}{\bibfnamefont{K.}~\bibnamefont{Lukas}},
	\bibinfo{author}{\bibfnamefont{B.}~\bibnamefont{Yu}},
	\bibinfo{author}{\bibfnamefont{S.}~\bibnamefont{Chen}},
	\bibinfo{author}{\bibfnamefont{C.}~\bibnamefont{Opeil}},
	\bibinfo{author}{\bibfnamefont{D.}~\bibnamefont{Broido}},
	\bibinfo{author}{\bibfnamefont{G.}~\bibnamefont{Chen}}, \bibnamefont{and}
	\bibinfo{author}{\bibfnamefont{Z.}~\bibnamefont{Ren}}, \bibinfo{journal}{J.
		Am. Chem. Soc.} \textbf{\bibinfo{volume}{134}}, \bibinfo{pages}{10031}
	(\bibinfo{year}{2012}).
	
	\bibitem[{\citenamefont{Pei et~al.}(2011{\natexlab{b}})\citenamefont{Pei, May,
			and Snyder}}]{Pei2011a}
	\bibinfo{author}{\bibfnamefont{Y.}~\bibnamefont{Pei}},
	\bibinfo{author}{\bibfnamefont{A.~F.} \bibnamefont{May}}, \bibnamefont{and}
	\bibinfo{author}{\bibfnamefont{G.~J.} \bibnamefont{Snyder}},
	\bibinfo{journal}{Adv. Energy Mater.} \textbf{\bibinfo{volume}{1}},
	\bibinfo{pages}{291} (\bibinfo{year}{2011}{\natexlab{b}}), ISSN
	\bibinfo{issn}{1614-6840}.
	
	\bibitem[{\citenamefont{Pei et~al.}(2014)\citenamefont{Pei, Gibbs, Gloskovskii,
			Balke, Zeier, and Snyder}}]{Pei2014}
	\bibinfo{author}{\bibfnamefont{Y.}~\bibnamefont{Pei}},
	\bibinfo{author}{\bibfnamefont{Z.~M.} \bibnamefont{Gibbs}},
	\bibinfo{author}{\bibfnamefont{A.}~\bibnamefont{Gloskovskii}},
	\bibinfo{author}{\bibfnamefont{B.}~\bibnamefont{Balke}},
	\bibinfo{author}{\bibfnamefont{W.~G.} \bibnamefont{Zeier}}, \bibnamefont{and}
	\bibinfo{author}{\bibfnamefont{G.~J.} \bibnamefont{Snyder}},
	\bibinfo{journal}{Adv. Energy Mater.} \textbf{\bibinfo{volume}{4}},
	\bibinfo{pages}{n/a} (\bibinfo{year}{2014}), ISSN \bibinfo{issn}{1614-6840}.
	
	\bibitem[{\citenamefont{Pei et~al.}(2012)\citenamefont{Pei, LaLonde, Heinz, and
			Snyder}}]{Pei2012}
	\bibinfo{author}{\bibfnamefont{Y.}~\bibnamefont{Pei}},
	\bibinfo{author}{\bibfnamefont{A.~D.} \bibnamefont{LaLonde}},
	\bibinfo{author}{\bibfnamefont{N.~A.} \bibnamefont{Heinz}}, \bibnamefont{and}
	\bibinfo{author}{\bibfnamefont{G.~J.} \bibnamefont{Snyder}},
	\bibinfo{journal}{Adv. Energy Mater.} \textbf{\bibinfo{volume}{2}},
	\bibinfo{pages}{670} (\bibinfo{year}{2012}), ISSN \bibinfo{issn}{1614-6840}.
	
	\bibitem[{\citenamefont{Wu et~al.}(2014)\citenamefont{Wu, Carrete, Zhang, Qu,
			Shen, Wang, Zhao, and He}}]{Wu2014}
	\bibinfo{author}{\bibfnamefont{H.}~\bibnamefont{Wu}},
	\bibinfo{author}{\bibfnamefont{J.}~\bibnamefont{Carrete}},
	\bibinfo{author}{\bibfnamefont{Z.}~\bibnamefont{Zhang}},
	\bibinfo{author}{\bibfnamefont{Y.}~\bibnamefont{Qu}},
	\bibinfo{author}{\bibfnamefont{X.}~\bibnamefont{Shen}},
	\bibinfo{author}{\bibfnamefont{Z.}~\bibnamefont{Wang}},
	\bibinfo{author}{\bibfnamefont{L.-D.} \bibnamefont{Zhao}}, \bibnamefont{and}
	\bibinfo{author}{\bibfnamefont{J.}~\bibnamefont{He}}, \bibinfo{journal}{NPG
		Asia Mater.} \textbf{\bibinfo{volume}{6}}, \bibinfo{pages}{e108}
	(\bibinfo{year}{2014}).
	
\end{thebibliography}

\end{document}